\documentclass[12pt,a4paper]{article}
\usepackage{amsfonts,latexsym}
\usepackage{amsmath,amssymb}
\usepackage{graphicx,color}

\renewcommand{\thefootnote}{\fnsymbol{footnote}}

\begin{document}

\vspace{12mm}
\begin{center}
{{{\Large {\bf Two  scalarizations of magnetically charged black holes }}}}\\[10mm]
{Yun Soo Myung\footnote{e-mail address: ysmyung@inje.ac.kr}}\\[8mm]

{Center for Quantum Spacetime, Sogang University, Seoul 04107, Republic of  Korea\\[0pt] }

\vspace{12mm}

\end{center}
 \begin{abstract}
    \noindent We study  two  scalarizations of  magnetically  charged black holes in the Einstein-Gauss-Bonnet-scalar theory with  the nonlinear electrodynamics (NED) term.
    For this purpose, two scalar coupling functions $f_1(\phi)$ and $f_2(\phi)$ are introduced to couple to Gauss-Bonnet (GB) and NED terms.
     The bald  black hole is described by
    mass ($M$) and action parameter ($\alpha$) with magnetic charge ($P$), which  becomes  the  quantum Oppenheimer-Snyder black hole for $P=M$ whose action is still unknown.
     We derive Smarr formula which describes   a correct thermodynamics  for the charged black hole.
    Two Davies points of heat capacity are not  identified with two critical onset mass and action  parameter for GB$^-$  scalarization with $f_1(\phi)$.
      Furthermore, the shadow radius analysis of  charged black hole is performed to include naked singularities and it is compared to the EHT observation.
      There is no constraints on the action parameter, whereas  new mass constraints are found.
      Finally, the NED$^+$ spontaneous scalarization with $f_2(\phi)$ leads to infinite branches of scalarized charged black holes.
\end{abstract}

\vspace{1.5cm}

\hspace{11.5cm}{Typeset Using \LaTeX}
\newpage
\renewcommand{\thefootnote}{\arabic{footnote}}
\setcounter{footnote}{0}

\vspace{2mm}


\section{Introduction}
\label{sec: Intro}
No-hair theorem states that a black hole can be completely characterized by only three observable classical parameters of  mass, electric charge, and angular momentum in Einstein gravity
\cite{Carter:1971zc,Ruffini:1971bza}. If a scalar field is minimally coupled to  gravitational and Maxwell  fields with a positive scalar potential, there is no scalar hair~\cite{Herdeiro:2015waa}.
However, its evasion (scalarized black holes) appeared  in the context of scalar-tensor theories possessing the nonminimal scalar coupling to either Gauss-Bonnet (GB) term~\cite{Doneva:2017bvd,Silva:2017uqg,Antoniou:2017acq}, or to Maxwell (M) term~\cite{Herdeiro:2018wub,Myung:2018vug,Myung:2018jvi}. The former is called curvature-induced (GB$^+$) spontaneous scalarization, while the later is called charged-induced (M$^+$) spontaneous scalarization because a positive (+) coupling constant was employed. In this case, one obtained infinite branches ($n=0,1,\cdots$) of scalarized (charged) black holes triggered by tachyonic instability.

On the other hand, the  spin-induced (GB$^-$) scalarization
of Kerr black holes described by mass $M$ and rotation parameter $a$  were studied  in the Einstein-Gauss-Bonnet-scalar (EGBS)
theory with a negative ($-$)  coupling constant~\cite{Cunha:2019dwb,Collodel:2019kkx}.
In this case, an $a$-bound of $a \ge a_c=0.5M$ was found to represent the condition for  onset of GB$^-$ scalarization~\cite{Dima:2020yac}. Here,  the critical rotation parameter $a_c$ was  computed in~\cite{Hod:2020jjy,Zhang:2020pko,Doneva:2020nbb}.  This implies  that  GB$^-$ scalarization is hard to  occur for the low rotation of $a<a_c$.
Later on, the spin-induced scalarized black holes have been numerically
constructed for high rotation and negative coupling constant in the EGBS theory~\cite{Herdeiro:2020wei,Berti:2020kgk}.
Similarly, the condition for spin-induced (M$^-$) scalarization of Kerr-Newman black holes described by mass ($M$), charge ($Q$), and rotation ($a$) was found to be  $a\ge a_c=0.672M$ with charge $Q=0.4$~\cite{Lai:2022spn} and the spin ($a$)-charge ($Q$) induced scalarization of Kerr-Newman black holes was studied in Einstein-Maxwell-scalar (EMS) theory with a negative coupling constant~\cite{Lai:2022ppn}.
In this case, one usually obtained single  branch of scalarized (rotating/charged) black holes.
However, this did not happen for Kerr black holes when the scalar is  nonminimally coupled to  the parity-violating  Chern-Simons term~\cite{Myung:2020etf}.

Interestingly,  the quantum Oppenheimer-Snyder (qOS) black hole was recently  found from studying the qOS gravitational collapse within the loop quantum cosmology~\cite{Lewandowski:2022zce}.
However, this approach has a handicap such that  one does  not know  its explicit action form $\mathcal{L}_{\rm qOS}$ to give the qOS black hole solution directly.
Even though various applications of this model were explored, we wish to focus on  thermodynamics~\cite{Dong:2024hod}, shadow radius analysis~\cite{Ye:2023qks}, and its GB$^-$ scalarization with quantum parameter ($\alpha$) which plays a role of the rotation parameter $a$ in Kerr black holes~\cite{Chen:2025wze}.

More recently, quantum parameter ($\alpha$)-mass ($M$) induced (GB$^-$) scalarization was found  in the EGBS theory with the unknown qOS action~\cite{Myung:2025pmx}: the conditions for GB$^-$ scalarization are given by  $\alpha_c(=1.2835)\le \alpha\le \alpha_e(=1.6875)$ for $M=1$ and $M_{\rm rem}(=0.7698)\le M \le M_c(=0.8827)$ for $\alpha=1$.  Importantly, we confirmed  that two Davies quantities ($\alpha_D,~M_D$) of heat capacity representing a sharp phase transition are equal to  two critical onset quantum parameter ($\alpha_c$) and mass ($M_c$)  for GB$^-$ scalarization, indicating  the first model to show a strong connection  between thermodynamics of bald black hole and onset of GB$^-$ scalarization clearly.
So, one expects  to find another models to show the strong connection  between thermodynamics of bald black hole and onset of  GB$^-$ scalarization.

In this direction, we are aware recently  that a candidate  for the qOS action was proposed  by introducing  the nonlinear electrodynamics (NED) term and the qOS black hole solution is recovered  upon imposing  $P=M$~\cite{Mazharimousavi:2025lld}. Inspired by this work, we will investigate the magnetically charged black holes  obtained from the EGBS theory with the NED term.

In the onset of this work, we wish to  study thermodynamics of the charged black hole described by mass ($M$) and action parameter ($\alpha$) with a fixed   magnetic charge ($P=0.6$).
We investigate  thermodynamic aspects of this black hole  by obtaining  thermodynamic quantities, checking  the first law thermodynamics and the Smarr formula, and mentioning  a sharp phase transition of heat capacity at Davies point.
The shadow radius analysis of the  charged black hole will be discussed to know how action parameter $\alpha$ is distinguished  from  mass $M$   when comparing them  with the EHT observation.
We check  that two Davies quantities ($\alpha_D,~M_D$) of heat capacity are not identified with   two critical onset action  parameter ($\alpha_c$) and mass ($M_c$)  for GB$^-$ scalarization obtained by introducing one coupling function $f_1(\phi)$ to GB term.  This implies that the charged black hole cannot not be regarded as  the qOS black hole.

Finally, we study  spontaneous scalarization by introducing the other coupling function $f_2(\phi)$ to the NED term to explore NED$^+$ scalarization which provides infinite branches of scalarized charged black holes.


\section{Magnetically charged black holes}
We introduce the Einstein-Gauss-Bonnet-scalar theory with the nonlinear electrodynamics (NED) as
\begin{equation}
\mathcal{L}_{\rm EGBSN}=\frac{1}{16 \pi}\Big[ R-2\partial_\mu \phi \partial^\mu \phi+ f_1(\phi) {\cal R}^2_{\rm GB}+ f_2(\phi){\cal L}_{\rm NED}\Big],\label{Action1}
\end{equation}
where
\begin{equation} \label{NED}
\mathcal{L}_{\rm NED}=-2\xi(\mathcal{F})^{\frac{3}{2}}
\end{equation}
with the Maxwell term $\mathcal{F}=F_{\mu\nu}F^{\mu\nu}$.
Here $\phi$ is the scalar field and ${\cal R}^2_{\rm GB}$ is the GB term defined by
\begin{equation}
{\cal R}^2_{\rm GB}=R^2-4R_{\mu\nu}R^{\mu\nu}+R_{\mu\nu\rho\sigma}R^{\mu\nu\rho\sigma}.\label{Action2}
\end{equation}
We choose a quadratic coupling $f_1(\phi)= 2\lambda \phi^2$ and $f_2(\phi)=1-\frac{\lambda \phi^2}{3}$ with a coupling constant $\lambda$ for exploring two scalarizations.

From the action (\ref{Action1}), we derive  the Einstein  equation
\begin{eqnarray}
 G_{\mu\nu}=2\partial _\mu \phi\partial _\nu \phi -(\partial \phi)^2g_{\mu\nu}+\Gamma_{\mu\nu}+T^{\rm NED}_{\mu\nu}, \label{equa1}
\end{eqnarray}
where $G_{\mu\nu}=R_{\mu\nu}-(R/2)g_{\mu\nu}$ is  the Einstein tensor and  $\Gamma_{\mu\nu}$   is given by
\begin{eqnarray}
\Gamma_{\mu\nu}&=&2R\nabla_{(\mu} \Psi_{\nu)}+4\nabla^\alpha \Psi_\alpha G_{\mu\nu}- 8R_{(\mu|\alpha|}\nabla^\alpha \Psi_{\nu)} \nonumber \\
&+&4 R^{\alpha\beta}\nabla_\alpha\Psi_\beta g_{\mu\nu}
-4R^{\beta}_{~\mu\alpha\nu}\nabla^\alpha\Psi_\beta  \label{equa2}
\end{eqnarray}
with
\begin{equation}
\Psi_{\mu}\equiv\frac{df_1(\phi)}{d\phi} \partial_\mu \phi= f'_1(\phi)\partial_\mu \phi.
\end{equation}
Here, the energy-momentum tensor takes the form
\begin{equation}
T^{\rm NED}_{\mu\nu}=4\xi f_1(\phi)\Big[\frac{3}{2}\sqrt{\mathcal{F}}F_{\mu\lambda}F^{\nu\lambda}-\frac{1}{4}\mathcal{F}^{3/2}g_{\mu\nu}\Big].
\end{equation}
The nonlinear Maxwell equation is given by~\cite{Mazharimousavi:2025lld}
\begin{equation}
\frac{1}{\sqrt{-g}}\partial_\mu\Big(\sqrt{-g}f_1(\phi)\sqrt{\mathcal{F}} F^{\mu\nu}\Big)=0.
\end{equation}
Choosing a magnetic field strength $F_{\theta\varphi}=q \sin \theta~(\mathcal{F}=2q^2/r^4)$ and  $\xi=2^{-3/2} 3\alpha /P$,  its energy-momentum tensor is determined  by
\begin{equation}
T^{\rm NED,\nu}_\mu=\frac{3\alpha P^2 f_1(\phi)}{r^6}{\rm diag}[-1,-1,2,2].
\end{equation}

On the other hand, the scalar field equation takes the form
\begin{equation}
\square \phi +\frac{1}{4}f'_1(\phi) {\cal R}^2_{\rm GB}- \frac{\xi}{2} f'_2(\phi)(\mathcal{F})^{\frac{3}{2}}=0 \label{s-equa}.
\end{equation}
Solving $G_{\mu\nu}=T^{\rm NED}_{\mu\nu}$ without couplings [$\phi=0$, $f'_1(\phi=0)=0$ and $f'_2(\phi=0)=0$], one finds a spherically symmetric  charged black hole  solution
\begin{equation} \label{ansatz}
ds^2_{\rm NED}= \bar{g}_{\mu\nu}dx^\mu dx^\nu=-g(r)dt^2+\frac{dr^2}{g(r)}+r^2d\Omega^2_2
\end{equation}
whose metric function is given by~\cite{Lewandowski:2022zce}
\begin{equation}
g(r)=1-\frac{2M}{r}+\frac{\alpha P^2}{r^4} \label{g-sol}
\end{equation}
with the action  parameter $\alpha$ having length dimension two.
We note  that Eq.(\ref{ansatz})  indicates  the charged black hole solution without scalar hair.
Selecting   $P=M$ by hand leads to the qOS black hole~\cite{Lewandowski:2022zce}, but it is not an extremal black hole.

\section{Thermodynamics for  charged black holes}
In this section, we investigate the thermodynamics of charged black hole with Smarr formula  in the grand canonical ensemble.
\begin{figure}
\centering
\mbox{
(a)
\includegraphics[angle =0,scale=0.4]{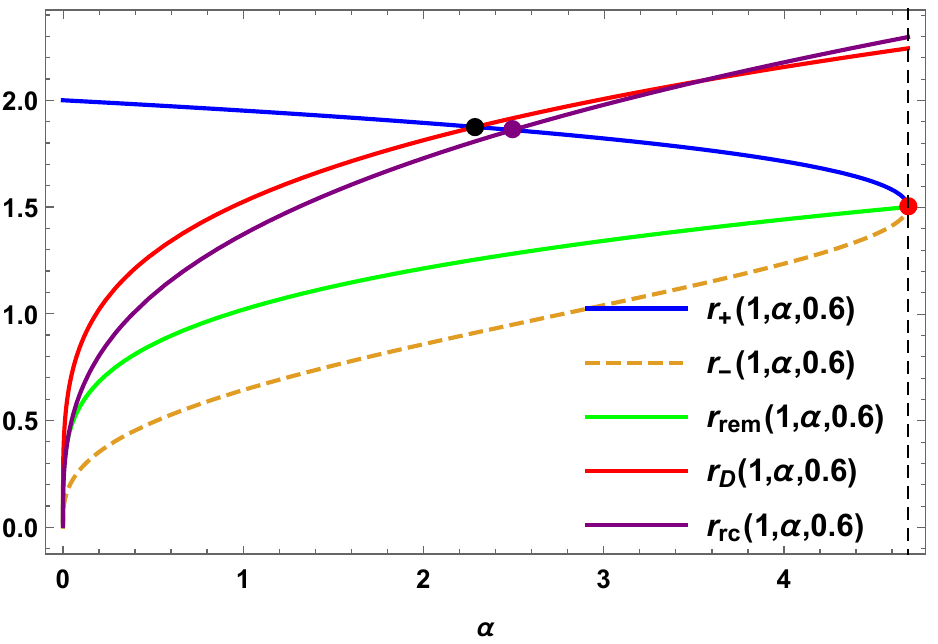}
(b)
\includegraphics[angle =0,scale=0.4]{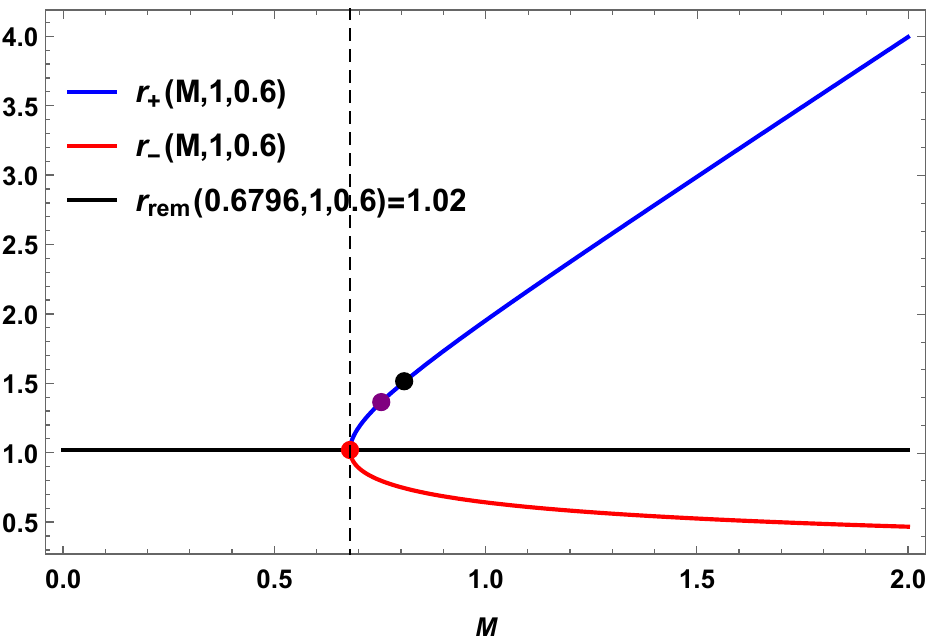}
}
\caption{ (a) Two outer/inner horizons $r_\pm(M=1,\alpha,P=0.6)$ are  functions of $\alpha\in[0,4.6875]$.  The remnant radius  $r_{\rm rem}(1,\alpha,0.6)$ as function of $\alpha$ includes  the extremal point at [(4.6875,1.5), red dot].  The Davies radius $r_D(1,\alpha,0.6)$ involves the Davies point at [(2.29,1.88), black dot], while the resonance radius $r_{\rm rc}(1,\alpha,0.6)$ implies a resonance point at [(2.49,1.86), purple dot].  (b) Two  horizons  $r_{\pm}(M,\alpha=1,P=0.6)$ are   function of   $M\in[0.6796,\infty]$, showing the lower bound (remnant point, red dot) for the mass of charged black hole with $r_{\rm rem}(0.6796,1,0.6)=1.02$. The black dot represents the Davies point at (0.82,1.52) while the purple dot denotes the resonance point at (0.76,1.37).}
\label{plot_modes}
\end{figure}
From $g(r)=0$, one finds two real solutions and two complex solutions
\begin{eqnarray}
&&r_i(M,\alpha,P),~{\rm for}~i=4,3,2,1, \label{f-roots}
\end{eqnarray}
where $r_4(M,\alpha,P)\to r_+(M,\alpha,P)$, $r_3(M,\alpha,P)\to r_-(M,\alpha,P)$, and $r_{2/1}$ become complex solutions.
The extremal black hole and the remnant  for $r_+=r_-$ occur for
\begin{equation}
\alpha_e(M,P)=\frac{27 M^4}{16 P^2},\quad M_{\rm rem}(\alpha,P)=2\sqrt{\frac{P\sqrt{\alpha}}{3\sqrt{3}}},
\end{equation}
which leads to $\alpha_e=4.6875$ for $M=1,P=0.6$ and $M_{\rm rem}=0.6796$ for $\alpha=1,P=0.6$.
As is shown in Fig. 1, there is an upper bound on $\alpha=\alpha_e$ as an extremal point for $M=1$, while there is a lower bound for the mass of  black hole (remnant mass $M=M_{\rm rem}$ for $\alpha=1$) but there is no  upper bound on the mass.  Hence, for $\alpha>\alpha_e$, the spacetime indicates naked singularity (NS), while for $M<M_{\rm rem}$, no horizon forms and the spacetime exhibits NS. These naked singularities could be arisen from strong gravitational lensing in section 4 when computing shadow radius analysis.

We find that black hole mass $m(M,\alpha,P)$ obtained from $g(r_+)=0$  after replacing $M$ with $m$, area-law entropy $S=\pi r_+^2$, the Hawking temperature defined by $T=\frac{\partial m}{\partial S}$,  heat capacity $C=\frac{\partial m}{\partial r_+}(\frac{\partial T}{\partial r_+})^{-1}$, and chemical potential $W_{\alpha}=\frac{\partial m}{\partial \alpha}$ as
\begin{eqnarray}
M&\to& m(M,\alpha,P)=\frac{\alpha P^2+r_+^4(M,\alpha,P)}{2r_+^3(M,\alpha,P)}, \label{ther1} \\
T(M,\alpha,P)&=&\frac{-3\alpha P^2+r_+^4(M,\alpha,P)}{4\pi r_+^5(M,\alpha,P)},  \label{ther2} \\
C(M,\alpha,P)&=-&\frac{2\pi r_+^2(M,\alpha,P)[r_+^4(M,\alpha,P)- 3\alpha P^2]}{ r_+^4(M,\alpha,P)-15 \alpha P^2},  \label{ther3} \\
W_\alpha(M,\alpha,P)&=&\frac{ P^2}{r_+^3(M,\alpha,P)}, \label{ther4}
\end{eqnarray}
where one checks that $T(M,\alpha,P)=T_\kappa(M,\alpha,P)$ obtained from the surface gravity $g'(r_+)/4\pi$.
\begin{figure*}[t!]
   \centering
  \includegraphics[width=0.5\textwidth]{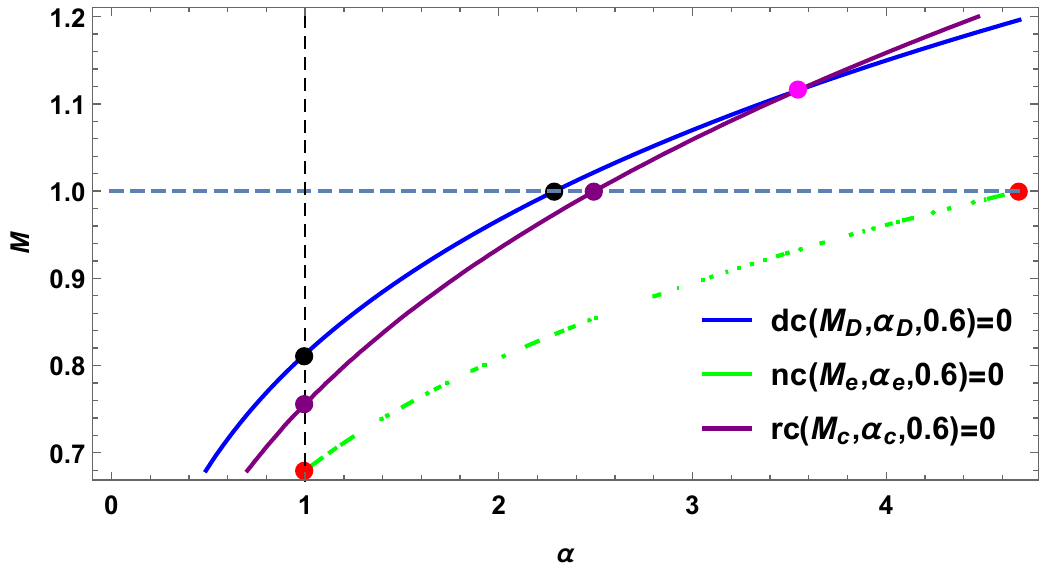}
\caption{ Three curves for $nc(M_e,\alpha_e,0.6)=0$, $dc(M_D,\alpha_D,0.6)=0$, and $rc(M_c,\alpha_c,0.6)=0$ for $M\in[0.6796,1.2]$ and $\alpha\in[0,4.6875]$ which include  extremal/remnant points (red dot), two Davies points (black dot), and two resonance  points (purple dot) when $M=1$ and $\alpha=1$. In general, one finds that $dc(M_D,\alpha_D,0.6)=0 \not=rc(M_c,\alpha_c,0.6)=0$, implying that Davies curve is not the same as resonance (critical onset) curve. However, a magenta dot at (3.54,1.12)  implies that  Davies points are equal to critical onset points.    }
\end{figure*}
The first law and Smarr formula are satisfied  as
\begin{equation}
dm=TdS+W_\alpha d\alpha,\quad m=2TS+4W_\alpha \alpha,
\end{equation}
which shows that the action parameter $\alpha$ plays the role of a thermodynamic variable, instead of  magnetic charge $P$.

From observing Eqs.(\ref{ther2}) and (\ref{ther3}), $T(M,\alpha,P)$ and $C(M,\alpha,P)$ are zero when their numerators are zero, showing extremal and remnant points,  while $C(M,\alpha,P)$ blows up when its denominator is zero, leading to the Davies point for a sharp phase transition. Here, we define the remnant and Davies curves [($r_{\rm rem}(M,\alpha,P),~r_{ D}(M,\alpha,P)$), see Fig. 1(a)] from the numerator zero condition  [$nc(M,\alpha,P)=0$] and the condition for denominator zero [$dc(M,\alpha,P)=0$] of $C$ as
\begin{eqnarray}
 nc(M,\alpha,P)&\equiv&r_+^4(M,\alpha,P)- 3\alpha P^2=0\to  r_{\rm rem}(M,\alpha,P)= (3\alpha P^2)^{1/4},  \label{rem-c} \\
 dc(M,\alpha,P)&\equiv&r_+^4(M,\alpha,P)- 15\alpha P^2=0\to  r_{D}(M,\alpha,P) = (15\alpha P^2)^{1/4},  \label{d-c}
\end{eqnarray}
where the former contains the extremal/remnant points ($C\to 0$, termination) and the latter includes the Davies point, implying  $C\to \infty$ (a sharp phase transition).
We note that $r_{\rm rem}(M,\alpha,P)$ could be obtained from the condition of numerator zero for temperature $T$.

Fig. 2  shows that three curves of $nc(M_e,\alpha_e,0.6)=0$, $dc(M_D,\alpha_D,0.6)=0$, and  $rc(M_c,\alpha_c,0.6)=0$ for $\alpha\in[0,4.6875]$ and $M\in[0.6796,1.2]$, including  extremal/remnant points (red dot), two Davies points (black dot), and two critical onset points (purple dot)  when $M=1$ and $\alpha=1$.
In general, one finds that $dc(M_D,\alpha_D,0.6)=0 \not=rc(M_c,\alpha_c,0.6)=0$, implying that Davies curve is not the same as critical onset curve.
However, a magenta dot at $(\alpha=3.54,M=1.12)$  implies especially that  Davies points are equal to critical onset points.
\begin{figure}
\centering
\mbox{
(a)
\includegraphics[angle =0,scale=0.4]{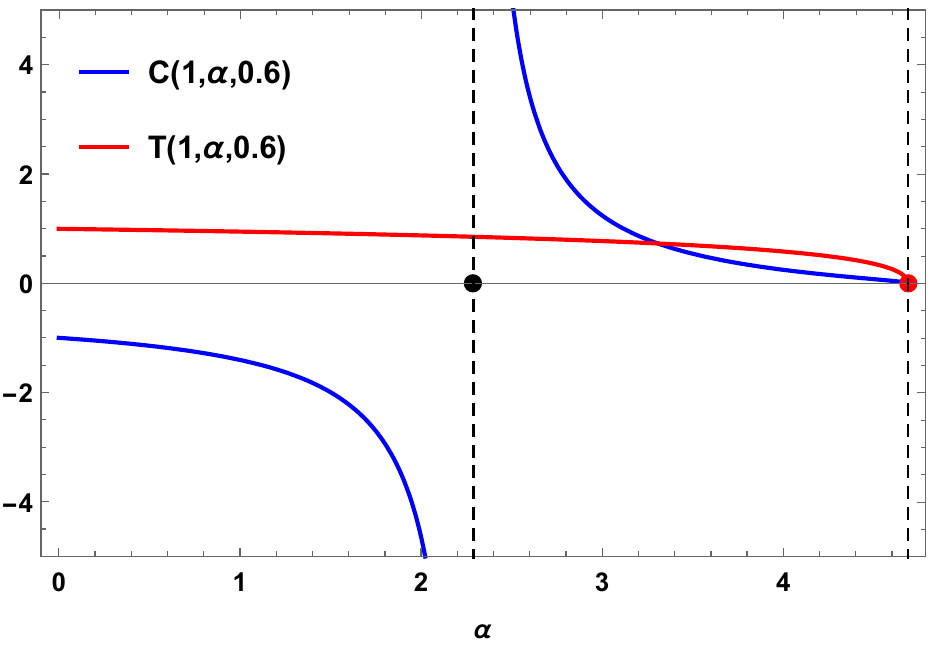}
(b)
\includegraphics[angle =0,scale=0.4]{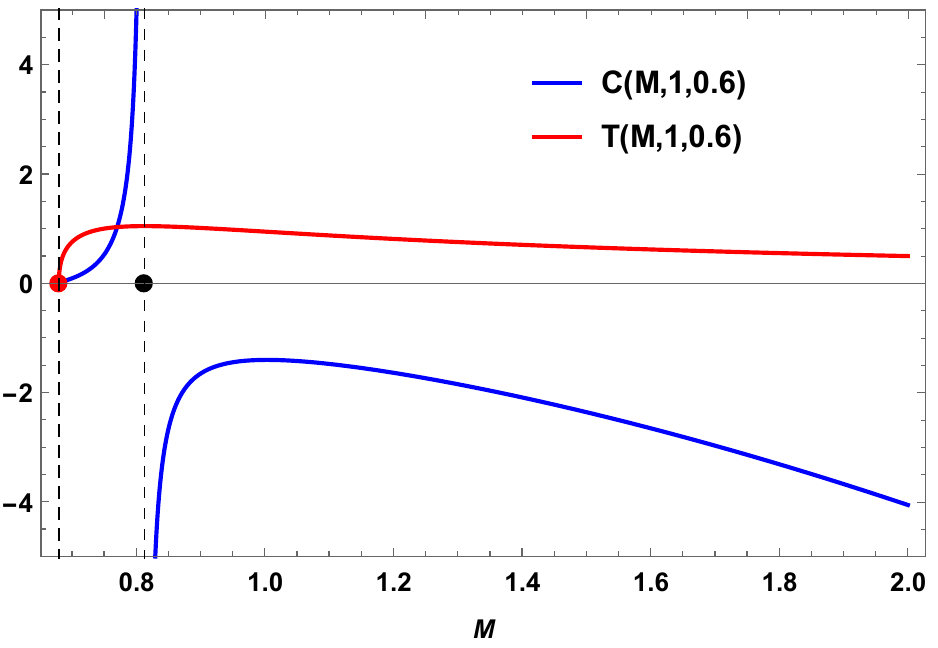}
}
\caption{ Heat capacity $C(M,\alpha,P)/|C_S(1,0,0)|$ with $|C_S(1,0,0)|=25.13$ and  temperature $T(M,\alpha,P)/0.04$  (a) Heat capacity $C(M=1,\alpha,P=0.6)$ blows up at Davies point ($\alpha_D=2.29,~\bullet$) and  are zero at the extremal point ($\alpha_e=4.6875,$ red dot). Temperature  $T(1,\alpha,0.6)$ is zero at extremal point. (b) Heat capacity $C(M,1,0.6)$ blows up at Davies point ($M_D=0.81,\bullet$) where the temperature $T(M,1,0.6)$ has the maximum. The heat capacity and temperature are zero at the remnant point ($M_{\rm rem}=0.6796,$ red dot). }
\label{plot_modes}
\end{figure}

We observe from Fig. 3 that the heat capacity $C(1,\alpha,0.6)/|C_S(1,0,0)|$ blows up at  Davies point ($\alpha_D=2.29$, $\bullet$).
Also, the  heat capacity  $C(M,1,0.6)/|C_S(1,0,0)|$ blows up at Davies point ($M_D=0.81$, $\bullet$) where the temperature $T(M,1,0.6)$ takes  the maximum value.

Finally, it is worth noting that the heat capacity and temperature are zero at the remnant point ($M_{\rm rem}=0.6796,$ red dot) as well as the extremal point ($\alpha_e=4.6875$). Even though the appearance of the remnant point was emphasized as an object to avoid singularity~\cite{Lewandowski:2022zce},  it is  regarded as an emerging (extremal)  point with $r_+=r_-$ in the thermodynamic viewpoint.
Otherwise, it may be a stable remnant after Hawking evaporation (starting from a finite size of charged black hole),  which  may resolve the information loss paradox~\cite{Dong:2024hod}.

\section{Shadow radius analysis}
We studied  thermodynamics of charged black hole to obtain  peculiar points and allowed regions for mass $M$ and action  parameter $\alpha$, whose peculiar points are extremal point, Davies point, and remnant point.
In this section, we need to analyze shadow radii by computing the  photon radius and critical impact parameter to extend the allowed regions of $M\in[0.6796,\infty$ ] and $\alpha\in[0,4.6875]$, whereas $P$ is fixed as 0.6 because its role is similar to $\alpha$ and all expressions come out as ``$\alpha P^2$".  The lower bound of $M$ and the upper bound of $\alpha$ can be extended to accommodate its NS (naked singularity)  by strong gravitational lensing.
We will  compare shadow radii with the recent EHT observation.

Requiring  the photon sphere, one finds two conditions with a geodesic potential $\tilde{V}(r)=g(r)/r^2$
\begin{equation} \label{cond-LR}
\tilde{V}(r=L)=\frac{1}{2b^2}, \quad \tilde{V}'(r=L)=0,
\end{equation}
where $b$ is  the critical impact parameter and $L$ represents the radius  of unstable photon sphere.
 Eq.(\ref{cond-LR}) implies   two relations
 \begin{equation}
 L^2=g(L)b^2,\quad 2g(L)=Lg'(L).
 \end{equation}
Here, the photon sphere radius and its critical impact parameter  are given by
\begin{eqnarray}
L(M,\alpha,P), \quad b(M,\alpha,P),\label{CI}
\end{eqnarray}
whose explicit forms are too complicated to write down here. It is important to  note that its $\alpha\in[4.6875,7.91]$-NS and $M\in[0.5963,0.6796]$-NS are arisen from the extensions of the photon sphere $L(M,\alpha,P)$.
This corresponds to  an effect of the strong gravitational lensing.
\begin{figure}
   \centering
   \mbox{
   (a)
  \includegraphics[width=0.4\textwidth]{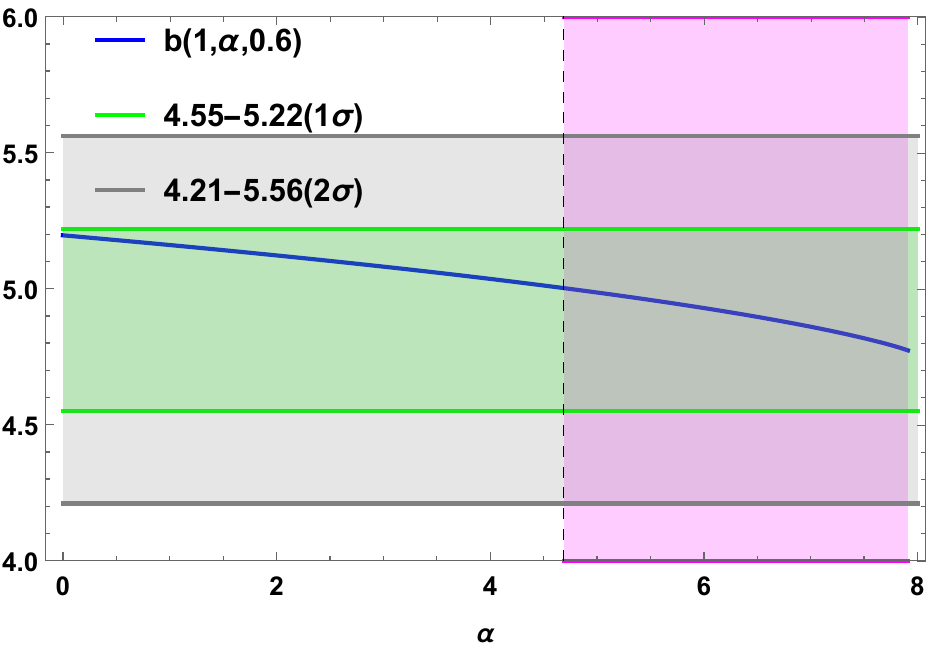}
 (b)
    \includegraphics[width=0.4\textwidth]{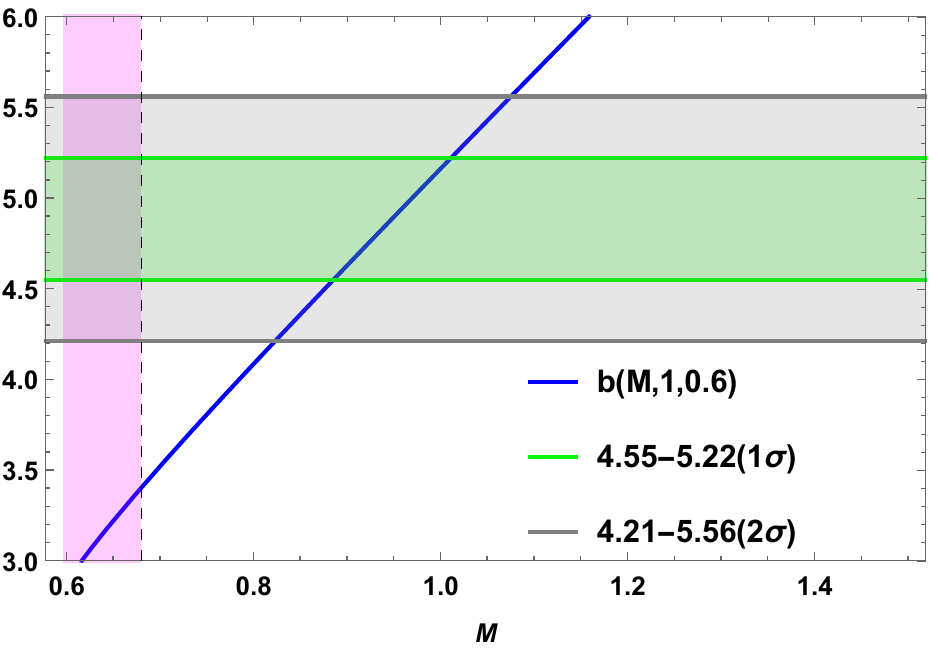}}
\caption{(a) Critical impact parameter  $b(M=1,\alpha,P=0.6)$ as a function of $\alpha\in[0,7.91]$.  There is one shaded region $\alpha\in$[4.6875,7.91] to represent  $\alpha$-NS region. A dashed line is at $\alpha_e=4.6875$ (extremal point).
 Here, we introduce $1\sigma$ and $2\sigma$ ranges from EHT observation.
 (b) A  critical impact parameter  $b(M,1,0.6)$ as a function of $M\in[0.5963,1.5]$.  There is one shaded region $M\in$[0.5963,0.6796] to denote $M$-NS region.  A dashed line is at the remnant point at $M_{\rm rem}=0.6796$.  }
\label{cip_modes}
\end{figure}

Considering  the EHT observation (Keck- and VLTI-based estimates for SgrA$*$~\cite{EventHorizonTelescope:2022wkp,EventHorizonTelescope:2022wok,EventHorizonTelescope:2022xqj}), the  $1\sigma$ constraint on the shadow radius $r_{\rm sh}=b$ indicates ~\cite{Vagnozzi:2022moj}
\begin{equation}
4.55\lesssim r_{\rm sh} \lesssim 5.22  \label{KV1}
\end{equation}
and the  $2\sigma$ constraint shows
\begin{equation}
4.21 \lesssim r_{\rm sh} \lesssim 5.56. \label{KV2}
\end{equation}
Let us observe  Fig. 4(a) for an explicit picture to test  with the EHT observation.
There is no constraints of the upper limit on its parameter  $\alpha$ for $M=1,P=0.6$. This means that  within $1\sigma$, the charged black hole  and its $\alpha$-NS   including an extremal point ($\alpha_e=4.6875$) are consistent with the EHT observation~\cite{Vagnozzi:2022moj}.
From  Fig. 4(b), we find that the charged black hole  has two constraints on the mass $M$ for $\alpha=1,P=0.6$: $0.886\lesssim M \lesssim 1.011 (1\sigma)$ and  $0.823\lesssim M \lesssim 1.076 (2\sigma)$ which belong to negative heat capacity.
However, its $M$-NS  is located beyond $2\sigma$ and belongs to positive heat capacity. This shows  a new constraint on mass $M$ found from the magnetically charged black holes.

\section{GB$^-$ scalarization with a negative $\lambda$}

Recently, it was shown that  the GB$^-$scalarization could  occur  when the
quantum  parameter satisfies $\alpha>\alpha_c=1.2835M^2$ in the EGBS theory with the unknown qOS action (\ref{Action1})~\cite{Chen:2025wze}.
In this section, we wish to investigate   the GB$^-$scalarization by computing critical onset mass $M_c$  and action parameter $\alpha_c$.
For this purpose, we need the scalar linearized equation which describes the scalar perturbation $\delta \phi$ propagating around (\ref{ansatz}).  It is derived by linearizing  Eq.(\ref{s-equa}) after choosing $f_1(\phi)=2\phi^2$ as
\begin{eqnarray}
  \left(\bar{\square}+ \lambda \bar{{\cal R}}^2_{\rm GB}\right)\delta \phi&=& 0. \label{scal-eq2}
\end{eqnarray}
Here the overbar  denotes computation based on the charged black hole background (\ref{ansatz}).
Importantly,  we note that ``$-\lambda \bar{{\cal R}}^2_{\rm GB}$" plays a role of
 an effective mass $m^2_{\rm eff}$  for $\delta \phi$.
Introducing a tortoise coordinate defined by $dr_*=dr/g(r)$ and considering
\begin{equation}
\delta\phi(t,r_*,\theta,\varphi)=\sum_m\sum^\infty_{l=|m|}\frac{\psi_{lm}(t,r_*)}{r}Y_{lm}(\theta,\varphi),
\end{equation}
Eq.(\ref{scal-eq2}) reduces  to the (1+1)-dimensional Klein-Gordon equation
\begin{equation} \label{mode-d}
\frac{\partial^2\psi_{lm}(t,r_*)}{\partial r_*^2} -\frac{\partial^2\psi_{lm}(t,r_*)}{\partial t^2}=V_1(r)\psi_{lm}(t,r_*),
\end{equation}
where the $s(l=0)$-mode potential $V_1(r)$ is given by
\begin{equation} \label{pot-c}
V_1(r)=g(r)\Big[\frac{2M}{r^3}-\frac{4\alpha P^2}{r^6}+m^2_{\rm eff}\Big]
\end{equation}
with its effective mass term
\begin{equation}
m^2_{\rm eff}=-\frac{48\lambda M^2}{r^{6}}\Big[\frac{3\alpha^2P^2}{r^6}-\frac{5\alpha P}{r^3} +1\Big].
\end{equation}
Eq.(\ref{mode-d}) is regarded as the (1+1)-dimensional mode decoupled equation  for $\psi_{lm}(t,r_*)$  because of a spherically symmetric background Eq.(\ref{ansatz}).
For $\lambda>0$ and  $\psi_{lm}(t,r_*)\sim u(r_*)e^{-i\omega t}$, one found  GB$^+$ scalarization of Schwarzschild black hole for large $\lambda$ and $\alpha=0$~\cite{Antoniou:2017acq,Doneva:2017bvd,Silva:2017uqg}.
For $\lambda<0$, one obtained spin-induced (GB$^-$) scalarization for Kerr black holes with rotation parameter $a$~\cite{Dima:2020yac,Herdeiro:2020wei,Berti:2020kgk,Lai:2022spn,Lai:2022ppn}. In this section, we consider the $\lambda<0$ case  and $\alpha\not=0$. In this case, one has to find the  critical onset parameter $\alpha_c$ and mass $M_c$ which determine the lower and upper bounds ($\alpha\ge \alpha_c$, $M\le M_c$) for the onset of scalarization  by using the Hod's approach~\cite{Hod:2020jjy}. Usually, the onset condition for GB$^-$ scalarization  is recovered  from analyzing the potential.

To get the critical onset parameters, we consider the potential term only
\begin{equation}
V_1(r)\psi_{lm}(t,r_*)=0.
\end{equation}
The onset of  scalarization is related closely to  an effective  binding potential well in the near-horizon whose two turning points of $r_{\rm in}$ and $r_{\rm out}$ are given  by the relation of $r_{\rm out}\ge r_{\rm in}=r_+(M,\alpha,P)$.  Two critical black holes with $\alpha=\alpha_c$ and $M=M_c$ denote the boundary between charged  and  scalarized charged black holes existing  in the limit of $\lambda \to -\infty$.
It is characterized by  the presence of a degenerate  binding potential well whose two turning points
merge at the outer horizon ($r_{\rm out}= r_{\rm in}=r_+$) as
\begin{eqnarray}
 m^2_{\rm eff}\psi_{lm}(t,r_*)=0, \quad {\rm for} \quad \alpha= \alpha_c,~M=M_c
\end{eqnarray}
in the limit of $\lambda  \to -\infty$.
At this stage, we introduce  the resonance condition which implies the resonance radius $r_{\rm rc}(M,\alpha,P)$ [see Fig. 1(a)] as
\begin{equation}
rc(M,\alpha,P)\equiv \frac{3\alpha^2P^2}{r^6_+(M,\alpha,P)}-\frac{5\alpha P}{r^3_+(M,\alpha,P)} +1=0\to r_{\rm rc}=\frac{(5\alpha P+\sqrt{13}\alpha P)^{1/3}}{2^{1/3}}.
\end{equation}
The critical onset (action)  parameter $\alpha_c$ is determined by the resonance condition because of $\psi_{lm}(t,r_*)\not=0$
\begin{eqnarray}
rc(M,\alpha_c,P)=0 \to  3\tilde{\alpha}_c^2-5\tilde{\alpha}_c+1 =0\label{res-con}
\end{eqnarray}
with $\tilde{\alpha}_c=\alpha_c P/r_+^3(M,\alpha_c,P)$.
Here we choose the small root
\begin{eqnarray}\label{critac}
\tilde{\alpha}_c=0.2324
\end{eqnarray}
for getting the critical onset parameter.
Solving $\tilde{\alpha}_c=\alpha_c/r_+^3(M,\alpha_c,P)=0.2324$ with $M=1$ and $P=0.6$,  one finds
the critical onset  parameter for  GB$^-$ scalarization as
\begin{eqnarray}
\alpha_c=2.4948.
\end{eqnarray}
Similarly, solving Eq.(\ref{res-con}) for $M_c$ with  $\alpha=1$ and $P=0.6$, we have
the critical onset mass for scalarization as
\begin{eqnarray}
 M_c=0.7556.
\end{eqnarray}
This implies that the charged black holes with either $\alpha \le \alpha_c$ or $M\ge M_c$ could not develop the tachyonic instability and thus, could not be realized as  scalarized charged black holes.
Here, we note that the other root of  $\tilde{\alpha}=1.4343$ to  Eq.~(\ref{res-con}) is not a physical solution for getting  $\alpha_c$ and $M_c$.

At this stage, it is important to note  that $(\alpha_c=2.4948,~M_c=0.7556)$ are not equal to Davies points $(\alpha_D=2.29,~M_D=0.81)$, which shows a clear difference to the equality  for qOS black hole~\cite{Myung:2025pmx}.
In general, one obtains  from Fig. 2 that $dc(M_D,\alpha_D,0.6)=0 \not=rc(M_c,\alpha_c,0.6)=0$, implying that Davies curve is  not equal to critical onset (resonance) curve.
However,  a magenta dot at $(\alpha=3.54,M=1.12)$  implies that  Davies points are equal to critical onset points but we do not understand why this point supports the equality between Davies and critical onset points.
\begin{figure*}[t!]
   \centering
    \mbox{
   (a)
  \includegraphics[width=0.4\textwidth]{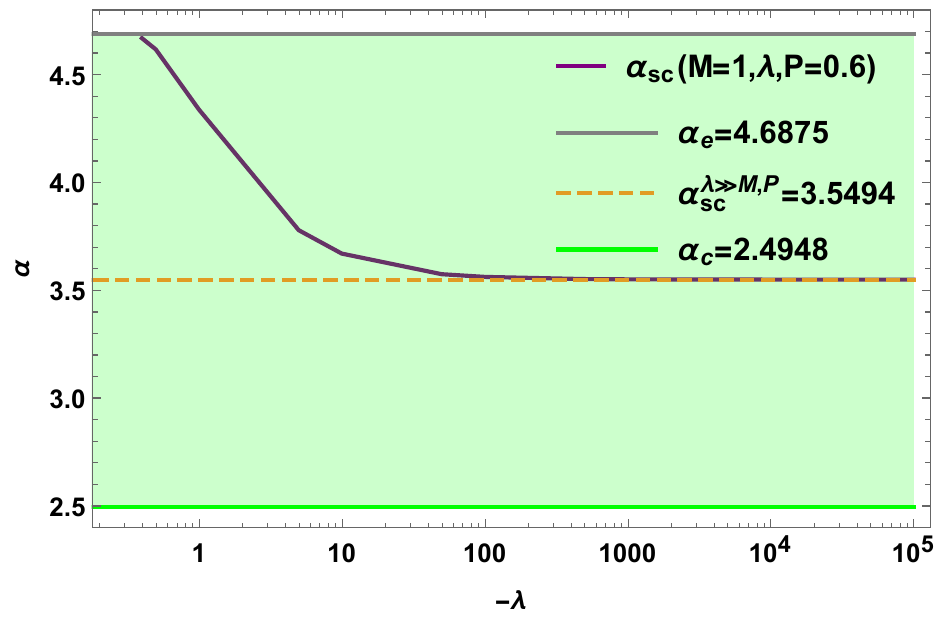}
 (b)
    \includegraphics[width=0.4\textwidth]{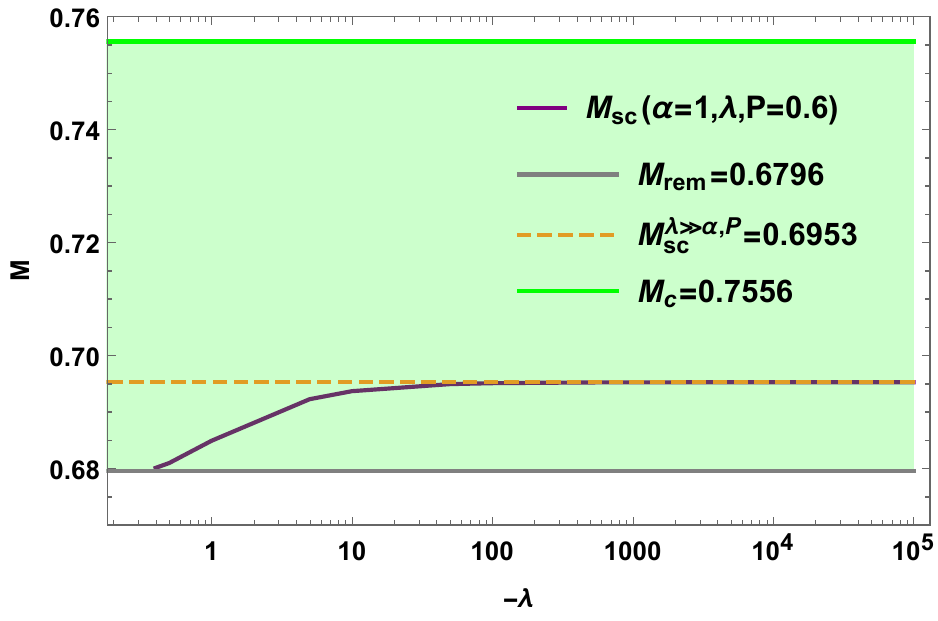}}
\caption{(a) Sufficient condition of $\alpha_{sc}(M=1,\lambda,P=0.6)$ with the critical onset parameter $\alpha=\alpha_c$ as the lower bound.  A dashed line denotes the sufficient condition $\alpha=\alpha_{sc}^{-\lambda\gg M,P}=3.5494$ for a large $-\lambda$. A top line denotes the extremal point at $\alpha=\alpha_e=4.6875$ as the upper bound.
(b) Graph for $M_{sc}(\alpha=1,\lambda,P=0.6)$ with  the critical onset mass $M=M_c$ as the upper bound  and the sufficient condition $M=M_{sc}^{-\lambda\gg  \alpha,P}=0.6953$ for a large $-\lambda$.  A bottom line represents the remnant point at $M=M_{\rm rem}=0.6796$ as the lower bound.}
\end{figure*}

To find the sufficient condition for the tachyonic instability, we use the  condition for instability suggested by Ref.\cite{Dotti:2004sh}
\begin{equation}
\int^\infty_{r_+(M,\alpha,P)}\Big[\frac{V_1(r,M,\alpha,P)}{g(r)}\Big] dr <0. \label{sc-ta}
\end{equation}
For large $-\lambda \gg M,\alpha,P$ where $M,\alpha,P$ are $\mathcal{O}(1)$,  this condition reduces to
\begin{equation}
\int^\infty_{r_+(M,\alpha,P)}m^2_{\rm eff} dr <0,
\end{equation}
which leads to
\begin{equation}
I(M,\alpha,P)\equiv \frac{6M^2[120 \alpha^2 P^2-275  \alpha P r_+^3(M,\alpha,P)+88r_+^6(M,\alpha,P)]}{55r_+^{11}(M,\alpha,P)}<0.
\end{equation}
This inequality  is not easy to solve for $\alpha$ because of the complexity of $r_+(M,\alpha,P)$.
Instead, we solve  $I(M,\alpha,P)=0$ to find two sufficient conditions  ($\alpha_{sc}^{-\lambda\gg M}=3.5494$ for $M=1,~M_{sc}^{-\lambda\gg  \alpha}=0.6953$ for $\alpha=1$) for large $-\lambda$.
As is shown in Fig. 5, the allowed regions for tachyonic instability
are depicted  as $\alpha_c(=2.4948) \le \alpha \le \alpha_e(=4.4875) $ [green shaded region] for $ M=1$  and $M_{\rm rem}(=0.6796)\le M\le M_c(=0.7556) $ [green shaded region] for $\alpha=1$. The lower region of $0<\alpha<\alpha_c$ and the upper region of $M>M_{c}$ are not acceptable   for GB$^-$ scalarization. It is worth noting that such restrictions imply single branch of scalarized charged black holes.

To find  $\alpha_{sc}(M=1,\lambda,P=0.6)$ and $M_{sc}(\alpha=1,\lambda,P=0.6)$, we analyze  the condition of Eq.(\ref{sc-ta}) numerically
for a given $\lambda$.
As is depicted in Fig. 4, we find that  $\alpha_{sc}(1,\lambda,0.6)$ is a decreasing function, connecting between  $\alpha_e=4.6875$ and  $\alpha_{sc}^{-\lambda\gg M,P}=3.5494$, while $M_{sc}(1,\lambda,0.6)$ is an increasing function, connecting between $M_{\rm rem}=0.6796$ and
$M_{sc}^{-\lambda\gg \alpha,P}=0.6953$.

 Finally, we would like to mention that to derive $\alpha_{th}(M=1,\lambda,P=0.6)$ connecting between $\alpha_e$ and $\alpha_c$  and $M_{th}(\alpha=1,\lambda,P=0.6)$ connecting between $M_{\rm rem}$ and $M_c$, one has to solve the Klein-Gordon equation (\ref{mode-d}) numerically with initial Gaussian wave packet~\cite{Chen:2025wze} because $V_1(r,M,\alpha,P)$ contains several terms. In this case, the fourth-order Runge-Kutta method plays a primary role in computing the time-domain profile of the scalar and the finite difference approach is introduced to validate the results. This is because the full onset for  GB$^-$ scalarization may show a complicated phenomena~\cite{Lai:2022spn,Lai:2022ppn}.

\section{NED$^+$ scalarization with a positive $\lambda$ }

In this section, we perform the NED$^+$ scalarization by choosing  $f_2(\phi)=1-\frac{\lambda \phi^2}{3}$ in Eq.(\ref{NED}).
The linearized scalar equation takes the from
\begin{equation}
(\bar{\square}-\tilde{m}^2_{\rm eff})\delta \phi=0. \label{linsca-2}
\end{equation}
This corresponds to the  spontaneous scalarization for the quadratic coupling ($\lambda>0$) to the NED.
Considering separation of variables
 \begin{equation}
 \delta \phi(t,r,\theta,\varphi) =\int \frac{ \psi(r)}{r} \sum_{lm} Y_{lm}(\theta) e^{i m \varphi} e^{-i\omega t} d\omega,
\end{equation}
the radial part of linearized scalar equation leads to
\begin{equation}
\frac{d^2 \psi(r)}{dr^2_*} +\Big[\omega^2-V_2(r)\Big] \psi(r) =0, \label{sch-eq1}
\end{equation}
where  the $s(l=0)$-mode scalar potential takes the form
\begin{equation} \label{pot-c}
V_2(r)=g(r)\Big[\frac{2M}{r^3}-\frac{4\alpha P^2}{r^6}+\tilde{m}^2_{\rm eff}\Big]
\end{equation}
with its single mass term
\begin{equation}
\tilde{m}^2_{\rm eff}=-\frac{\lambda \alpha P^2}{r^6}.
\end{equation}
This term induces onset of spontaneous scalarization through tachyonic instability.
\begin{figure*}[t!]
   \centering
    \mbox{
   (a)
  \includegraphics[width=0.4\textwidth]{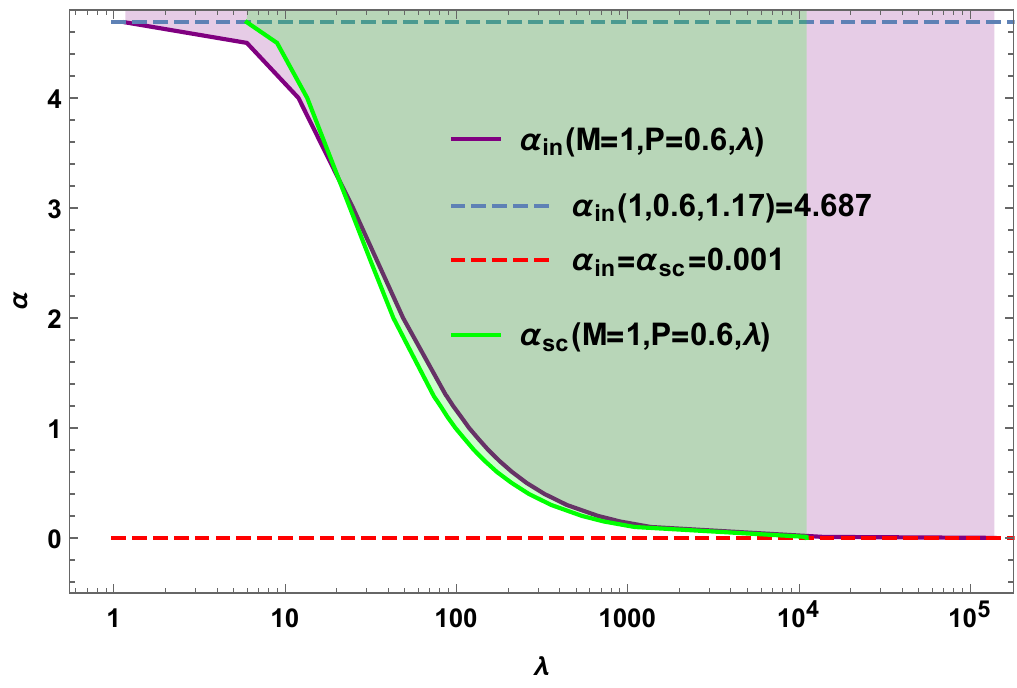}
 (b)
    \includegraphics[width=0.4\textwidth]{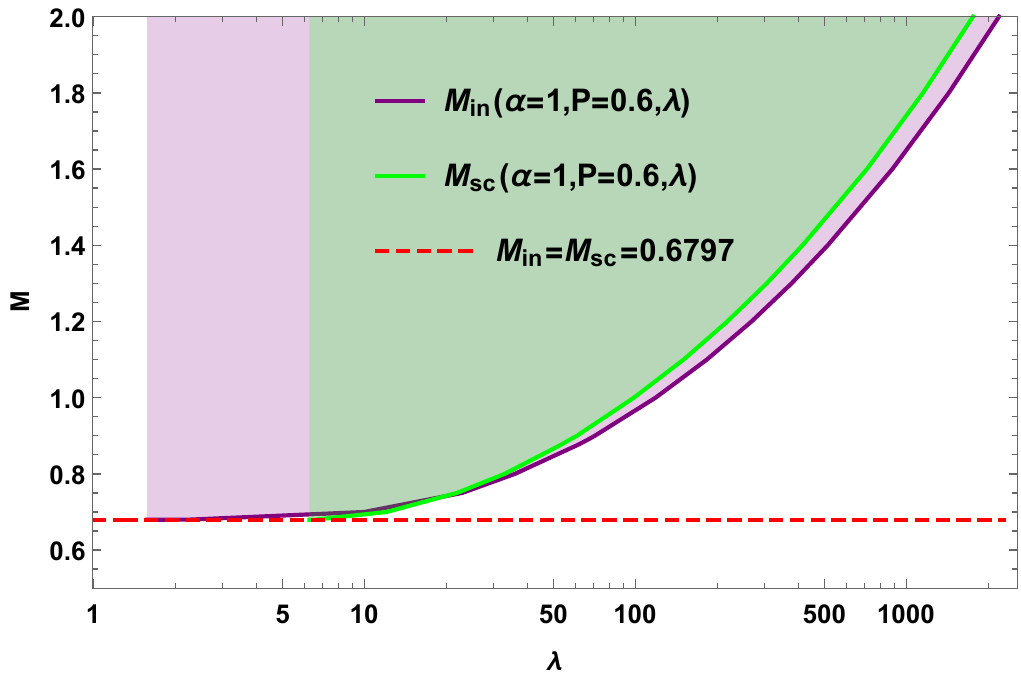}}
\caption{(a) Sufficient conditions of $\alpha_{sc}(M=1,P=0.6,\lambda)$ and  $\alpha_{in}(M=1,P=0.6,\lambda)$.  A bottom line denotes  $\alpha_{sc}=\alpha_{in}=0.001$,  while a top line represents  the extremal point at $\alpha=\alpha_e=4.6875$ as the upper bound.
(b) Graphs for $M_{sc}(\alpha=1,P=0.6,\lambda)$ and $M_{in}(\alpha=1,P=0.6,\lambda)$.   A bottom line represents the remnant point at $M_{sc}=M_{in}=M_{\rm rem}=0.6796$ as the lower bound.}
\end{figure*}

To find the sufficient conditions $\alpha\ge  \alpha_{sc}(M=1,P=0.6,\lambda)$ and $M\ge M_{sc}(\alpha=1,P=0.6,\lambda)$ for tachyonic instability, we analyze  the following condition:
\begin{equation}
\int^\infty_{r_+(M,\alpha,P)}\Big[\frac{V_2(r)}{g(r)}\Big] dr <0 \label{sc-pta}
\end{equation}
for a given $\lambda$ numerically.
As is depicted in Fig. 6, we find that  $\alpha_{sc}(1,0.6,\lambda)$ is a decreasing function of $\lambda$, connecting between   $\alpha_e=4.6875$ and $\alpha=0$, while $M_{sc}(1,0.6,\lambda)$ is an increasing function of $\lambda$, connecting between $M_{\rm rem}=0.6796$ and $M=\infty$. The shaded regions of $\alpha\ge \alpha_{sc}(1,0.6,\lambda)$ and $M\ge M_{sc}(1,0.6,\lambda)$  are sufficiently  unstable regions.
Here, we do not find  any critical onset parameters encountered in GB$^-$ scalarization, which means that allowed regions for action parameter $\alpha_{sc}$ and mass $M_{sc}$ are extended to include   $\alpha_{sc}\in[0,4.6875]$ and $M_{sc}\in[0.6796,\infty]$, compared to narrow ranges of  $\alpha_{sc}\in[3.5494,4.6875]$ and $M_{sc}\in[0.6796,0.6953]$ for GB$^-$ scalarization.

To obtain $\alpha_{\rm in}(M,P,\lambda)$, it is  proposed that  the spatially regular  scalar configurations (scalar clouds)  described by Eq.(\ref{linsca-2}) with $\omega=0$  could be investigated
 by employing  the standard WKB technique~\cite{Hod:2019ulh}.
A second-order WKB method  may be applied for obtaining the bound states of the scalar potential $V_2(r_*)$ approximately to yield the quantization condition
\begin{equation}
\int^{r_*^{\rm out}}_{r_*^{\rm in}}dr_* \sqrt{-V_2(r_*)}=\Big(n-\frac{1}{4}\Big)\pi,\quad n=1,2,3,\cdots, \label{wkb1}
\end{equation}
where $r_*^{\rm out}$ and $r_*^{\rm in}$ are radial turning points which satisfy  $V_2(r_*^{\rm out})=V_2(r_*^{\rm in})=0$.
We could express  Eq.(\ref{wkb1}) in terms of the radial coordinate  $r$ as
\begin{equation}
\int^{r_{\rm out}}_{r_{\rm in}}dr \frac{\sqrt{-V_2(r)}}{g(r)}=\Big(n-\frac{1}{4}\Big)\pi,\quad n=1,2,3,\cdots. \label{wkb2}
\end{equation}
Here, radial turning points $(r_{\rm out},r_{\rm in})$ are determined by the two conditions
\begin{equation}
g(r_{\rm in})=0,\quad 2Mr_{\rm out}^3-(\lambda+4)\alpha P^2=0,
\end{equation}
which imply
\begin{equation}
r_{\rm in}=r_+(M,\alpha,P),~\quad r_{\rm out}= \Big[ \frac{(\lambda+4)\alpha P^2}{2M}\Big]^{1/3}.
\end{equation}
For large $\lambda (r_{\rm out})$, the WKB integral (\ref{wkb2}) could be approximated by considering the last term in (\ref{pot-c}) as
\begin{equation}
\sqrt{\lambda}\cdot \sqrt{\alpha} P\int^{\infty}_{r_+} \frac{dr}{\sqrt{r^6 g(r)}}\equiv\sqrt{\lambda} I_n(M,\alpha,P) =\Big(n+\frac{3}{4}\Big)\pi,\quad  n=0,1,2,\cdots,
\end{equation}
which could be integrated numerically to yield
\begin{equation} \label{alphan}
\lambda_{in,n}(M,\alpha,P)=\frac{\pi^2(n+3/4)^2}{I^2_n(M,\alpha,P)},\quad  n=0,1,2,\cdots.
\end{equation}
\begin{figure*}[t!]
   \centering
  \includegraphics[width=0.5\textwidth]{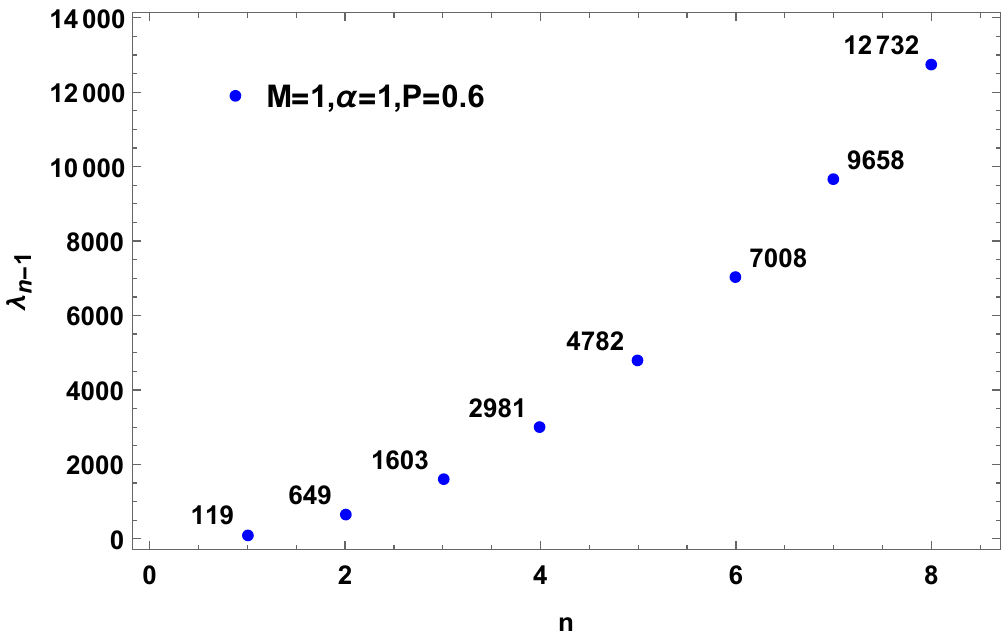}
\caption{(Left) Infinite  branches  $\lambda_{ in,n-1}(M=1,\alpha=1,P=0.6)$ are starting points  for $n=1,\cdots,8$. The fundamental branch ($n=0$) is allowed as $\lambda\in[\lambda_{in,0}=119,\infty]$ for $n=1$, while
the first excited ($n=1$) branch is allowed as $\lambda\in[\lambda_{ in,1}=649,\infty]$ for $n=2$. }
\end{figure*}
We display  $\alpha_{in}(M=1,P=0.6,\lambda_{in,0})\in[0,4.6875]$ and  $M_{in}(\alpha=1,P=0.6,\lambda_{in,0})\in[0.6796,\infty]$ in Fig. 6, but two allowed regions are given by  $\alpha_{in}\in[2.4948,4.6875]$ and $M_{in}\in[0.6796,0.7556]$ for GB$^-$ scalarization.
We note  that $\alpha_{in}(1,0.6,\lambda_{in,0}) [\simeq \alpha_{sc}(1,0.6,\lambda)]$ is a monotonically decreasing function of $\lambda$.
However, we observe  an inequality  for $\lambda<34$ carefully
\begin{equation}
\alpha_{ in}(1,0.6,\lambda )\le \alpha_{ th}(1,0.6,\lambda)\le \alpha_{sc}(1,0.6,\lambda), \label{ineq-R}
\end{equation}
while for $\lambda >34$, one finds the other inequality
\begin{equation}
\alpha_{ sc}(1,0.6,\lambda ) \le \alpha_{ th}(1,0.6,\lambda)\le \alpha_{in}(1,0.6,\lambda) \label{ineq-L}
\end{equation}
with the threshold of instability $\alpha_{ th}(1,0.6,\lambda)$ to be determined.

On the other hand,  $M_{in}(\alpha=1,P=0.6,\lambda_{in,0}) [\simeq M_{sc}(1,0.6,\lambda)]$ is a monotonically increasing  function of $\lambda$.
In this case, one  has an inequality  for $\lambda<35$
\begin{equation}
M_{ sc}(1,0.6,\lambda )\le M_{ th}(1,0.6,\lambda)\le M_{in}(1,0.6,\lambda), \label{Mineq-R}
\end{equation}
while for $\lambda >35$, one finds the other inequality
\begin{equation}
M_{ in}(1,0.6,\lambda )\le M_{ th}(1,0.6,\lambda)\le M_{sc}(1,0.6,\lambda) \label{Mineq-L}
\end{equation}
with the threshold of instability $M_{ th}(1,0.6,\lambda)$ for the mass to be determined.
Others of $\lambda_{in,n\not=0}(M,\alpha,P) $ may be  used to determine starting  points for infinite branches ($n=1,2,3,\cdots$)  (see Fig. 7).

 Finally, one has to solve Eq.(\ref{sch-eq1}) with $\omega\to i\Omega$  to determine $\alpha_{ th}(1,0.6,\lambda)$ and $M_{ th}(1,0.6,\lambda)$.

\section{Discussions}

We have obtained two scalarizations of magnetically charged black holes  in the EGBS theory with the NED term by considering two coupling functions $f_1(\phi)$ and $f_2(\phi)$.
An important feature of this study is that the charged black  hole was described  by mass $M$ and  action parameter $\alpha$ with a fixed  magnetic charge $P=0.6$.
This means that the magnetic charge $P$  plays a subsidiary role in studying charged black holes, contrasting with action parameter $\alpha$.
This reflects the nature of the present model as charged quantum Oppenheimer-Snyder model~\cite{Mazharimousavi:2025lld}  because its ($P$) role is similar to $\alpha$ and all expressions concerning $P$ came out as either ``$\alpha P^2$" or ``$\alpha P$".

We studied  thermodynamics of charged black holes to obtain  peculiar points and allowed regions for mass ($M\in[M_{\rm rem}=0.6796,\infty]$ for $\alpha=1,P=0.6$) and action  parameter ($\alpha\in[0,\alpha_e=4.6875]$ for $M=1,P=0.6$), whose peculiar points are extremal point ($\alpha_e$), Davies points ($\alpha_D,M_D$), and remnant point ($M_{\rm rem}$).
Thermodynamic study was done   by computing  thermodynamic quantities, checking  the first law thermodynamics and the Smarr formula, and considering  a sharp phase transition at Davies points ($\alpha_D=2.29$ for $M=1,P=0.6$ and $M_D=0.81$ for $\alpha=1,P=0.6$) related to the heat capacity.
Even though the presence of the remnant mass ($M_{\rm rem}$) was emphasized as an object to avoid singularity~\cite{Lewandowski:2022zce},  it has nothing but an emerging (extremal)  point with $r_+=r_-$ in the thermodynamic viewpoint.
Otherwise, it may be a stable remnant after Hawking evaporation (starting from a finite size of charged black hole),  which  may resolve the information loss paradox~\cite{Dong:2024hod}.
To this model, we did not find a close connection between thermodynamics of bald black hole and  onset of GB$^-$ scalarization
because  Davies curve ($\alpha_D,M_D$) from  heat capacity are not identified with  critical onset curve ($\alpha_c,M_c$) in  the EGBS theory with the NED action.
However, a magenta dot at $(\alpha=3.54,M=1.12$) was discovered such that  Davies points are equal to critical onset points.

Furthermore, the shadow radius analysis of  charged black holes was performed to include two naked singularities ($\alpha$-NS$\in[\alpha_e,7.91]$ and $M$-NS$\in[0.5963,M_{\rm rem}]$) by considering strong gravitational lensing (extension of allowed regions) and was  compared to the EHT observation.
We found that there is no constraints on the action parameter ($\alpha,\alpha_e,\alpha$-NS). New constraints are obtained for the mass ($M$) of charged black holes, but no constraint on $M$-NS is found.

For $f_1(\phi)$ to GB term (${\cal R}^2_{\rm GB}$) with $f_2(\phi)=0$ and $\lambda<0$, we found GB$^-$ scalarization with critical onset parameters ($\alpha_c=2.4948$ for $M=1,P=0.6$ and  $M_c=0.7556$ for $\alpha=1,P=0.6$), implying the single branch of scalarized charged black holes.
For $f_2(\phi)$ to NED term ($\mathcal{F}^2$) with  $f_1(\phi)=0$ and $\lambda>0$, we found NED$^+$ spontaneous scalarization, implying infinite  branches ($n=0,1,\cdots$) of scalarized charged black holes.
It was found that the allowed unstable regions are narrow as  $\alpha_{in}(M=1,P=0.6, \lambda)\in[\alpha_c,\alpha_e]$ and $M_{in}(\alpha=1,P=0.6,\lambda)\in[M_{\rm rem},M_c]$ for the single branch of  GB$^-$ scalarization, while they are wide as given by  $\alpha_{in}(M=1,P=0.6,\lambda)\in[0,\alpha_e]$ and $M_{in}(\alpha=1,P=0.6,\lambda)\in[M_{\rm rem},\infty]$ for the fundamental ($n=0$) branch of NED$^+$ scalarization.

Hence,  a further step would be  to perform the full scalarization with  action number ($\alpha$) and  mass ($M$)  to obtain  scalarized charged black holes  by solving full equations (\ref{equa1}) and (\ref{s-equa}).

\section{Acknowledgments}

Y.S.M. is supported by the National Research Foundation of Korea (NRF) grant
 funded by the Korea government (MSIT) (RS-2022-NR069013).

\end{document}